\documentstyle[12pt]{article}

\catcode`\@=11
\long\def\@makefntext#1{
\protect\noindent \hbox to 3.2pt {\hskip-.9pt  
$^{{\ninerm\@thefnmark}}$\hfil}#1\hfill}                

\def\@makefnmark{\hbox to 0pt{$^{\@thefnmark}$\hss}}  
        
\def\ps@myheadings{\let\@mkboth\@gobbletwo
\def\@oddhead{\hbox{}
\rightmark\hfil\ninerm\thepage}   
\def\@oddfoot{}\def\@evenhead{\ninerm\thepage\hfil
\leftmark\hbox{}}\def\@evenfoot{}
\def\sectionmark{}\def\subsectionmark{}}

\renewcommand{\thefootnote}{\fnsymbol{footnote}}

\renewcommand{\subsubsection}[1]
{\vspace*{0.6cm}\addtocounter{subsubsectionc}{1}
        \noindent
{\normalsize\rm\thesectionc.\thesubsectionc.\thesubsubsectionc. 
        #1}\par\vspace*{0.6cm}}

\newcounter{appendixc}
\newcounter{subappendixc}[appendixc]
\newcounter{subsubappendixc}[subappendixc]

\renewcommand{\appendix}[1] {\vspace*{0.6cm}
        \refstepcounter{appendixc}
        \setcounter{figure}{0}
        \setcounter{table}{0}
        \setcounter{equation}{0}
        \renewcommand{\thefigure}{\Alph{appendixc}.\arabic{figure}}
        \renewcommand{\thetable}{\Alph{appendixc}.\arabic{table}}
        \renewcommand{\theappendixc}{\Alph{appendixc}}
        \renewcommand{\theequation}{\Alph{appendixc}.\arabic{equation}}
        \noindent{\bf Appendix \theappendixc #1}\par\vspace*{0.4cm}}

\def\abstracts#1{{
       
\centering{\begin{minipage}{12.2truecm}\footnotesize\baselineskip=12pt\noindent
        \centerline{\footnotesize ABSTRACT}\vspace*{0.3cm}
        \parindent=0pt #1
        \end{minipage}}\par}} 

\newcounter{itemlistc}
\newcounter{romanlistc}
\newcounter{alphlistc}
\newcounter{arabiclistc}

\newcommand{\fcaption}[1]{
        \refstepcounter{figure}
        \setbox\@tempboxa = \hbox{\footnotesize Fig.~\thefigure. #1}
        \ifdim \wd\@tempboxa > 6in
           {\begin{center}
        \parbox{6in}{\footnotesize\baselineskip=12pt Fig.~\thefigure. #1}
            \end{center}}
        \else
             {\begin{center}
             {\footnotesize Fig.~\thefigure. #1}
              \end{center}}
        \fi}

\newcommand{\tcaption}[1]{
        \refstepcounter{table}
        \setbox\@tempboxa = \hbox{\footnotesize Table~\thetable. #1}
        \ifdim \wd\@tempboxa > 6in
           {\begin{center}
        \parbox{6in}{\footnotesize\baselineskip=12pt Table~\thetable. #1}
            \end{center}}
        \else
             {\begin{center}
             {\footnotesize Table~\thetable. #1}
              \end{center}}
        \fi}

\def\@citex[#1]#2{\if@filesw\immediate\write\@auxout
        {\string\citation{#2}}\fi
\def\@citea{}\@cite{\@for\@citeb:=#2\do
        {\@citea\def\@citea{,}\@ifundefined
        {b@\@citeb}{{\bf ?}\@warning
        {Citation `\@citeb' on page \thepage \space undefined}}
        {\csname b@\@citeb\endcsname}}}{#1}}

\newif\if@cghi
\def\cite{\@cghitrue\@ifnextchar [{\@tempswatrue
        \@citex}{\@tempswafalse\@citex[]}}
\def\citelow{\@cghifalse\@ifnextchar [{\@tempswatrue
        \@citex}{\@tempswafalse\@citex[]}}
\def\@cite#1#2{{$\null^{#1}$\if@tempswa\typeout
        {IJCGA warning: optional citation argument 
        ignored: `#2'} \fi}}

 1
 1
 1

\font\ninerm=cmr9

\begin{document}

\centerline{\normalsize\bf $1/Q^2$ TERMS AND LANDAU SINGULARITY\footnote
{Talk presented at the Moriond Conference ``QCD and High Energy Hadronic 
Interactions'', Les Arcs, France, March 22-29, 1997.}}
\baselineskip=22pt

\vspace{0.6cm}
\centerline{\footnotesize Georges GRUNBERG}

\baselineskip=13pt
\centerline{\footnotesize\it Centre de Physique Th\'eorique de l'Ecole 
Polytechnique\footnote{CNRS UPRA 0014}}
\baselineskip=12pt
\centerline{\footnotesize\it 91128 Palaiseau Cedex - France}
\vspace{0.6cm}

\vspace{0.9cm}
\abstracts{ Standard power-behaved contributions in QCD arising from 
non-perturbative effects at low scale can be 
described, as shown by Dokshitzer, Marchesini and Webber, with the notion of 
an infrared regular effective coupling. In their approach, a non-perturbative 
contribution to the coupling, essentially restricted to low scales, 
parametrizes the non-perturbative power corrections. I argue that their 
framework naturally allows for another type of power contributions, arising 
from short distances (hence unrelated to
 renormalons
and the operator product expansion)  which  
appear in the process of  removing the Landau singularity present in 
perturbation theory.   A natural definition of an infrared 
finite perturbative
coupling is suggested within the dispersive method. Implications for the tau 
hadronic width, where ${\cal O}(1/Q^2)$ contributions can
be generated,
 are pointed out. }
\vspace{7cm}
CPTh/PC 510.0597
\hspace{9cm}
May 1997

\vspace{0.5cm}hep-ph/yy mm nnn
\newpage
\pagestyle{plain} 
\normalsize\baselineskip=15pt
\setcounter{footnote}{0}
\renewcommand{\thefootnote}{\alph{footnote}}
The study of power corrections in QCD has been the subject 
of active
investigations in recent years. Their importance for a precise determination of 
$\alpha_s$ has
been recognized, and  various  techniques (renormalons, finite gluon 
mass, dispersive
approach) have been devised to deal with situations where the standard 
operator product
expansion (OPE) does not apply. In this
talk (which is a summary of$^{1)}$) , I  focuss on the dispersive 
approach$^{2)}$, based on the notion of an 
infrared (IR)
regular$^{3)}$   QCD coupling, where  a non-perturbative contribution 
to the coupling,
essentially restricted to low scales, parametrizes the  power corrections. 
I point out that
within this framework,  it is  very natural to expect the existence of new type 
of power
contributions {\em of ultraviolet (UV) origin, hence not controlled by the OPE},
 related to the
removal of the IR Landau singularity presumably present in the perturbative part
 of the coupling.

Consider the contribution to an Euclidean (quark dominated) observable arising 
from dressed
single gluon virtual exchange, which takes the generic form (after subtraction 
of the Born term):
\begin{equation}D(Q^2) =\int_{0}^\infty{dk^2\over k^2}\ \alpha_s(k^2)\ 
\varphi\left(k^2\over
Q^2\right)\end{equation}
The ``physical'' coupling $\alpha_s(k^2)$ is assumed to be IR regular, and thus 
must differ from the perturbative coupling  $\alpha_s^{PT}(k^2)$ by a ``power 
correction'' piece $\delta\alpha_s(k^2)$. To determine the various types of
power 
contributions, it is appropriate to disentangle long from short distances ``a la
 SVZ'' with an IR cutoff $\Lambda_I$:
\begin{equation}D(Q^2) = \int_0^\infty{dk^2\over k^2}\ \alpha_s^{PT}(k^2)\ 
\varphi\left(k^2
\over Q^2\right) +
\int_{0}^{\Lambda_I^2}{dk^2\over k^2}\ \delta\alpha_s(k^2)\ \varphi\left(k^2
\over Q^2\right) +
\int_{\Lambda_I^2}^\infty{dk^2\over k^2}\ \delta\alpha_s(k^2)\ \varphi\left(k^2
\over
Q^2\right)\end{equation}
The first integral on the right hand side of eq.(2) may be identified to the 
Borel sum $D_{PT}(Q^2)$ of perturbation theory. The second integral gives 
``long distance ''
power corrections which correspond to the standard OPE
``condensates''$^{4)}$. If the Feynman diagram kernel $\varphi
\left(k^2\over Q^2\right)$ is ${\cal O}\left((k^2/Q^2)^n\right)$ at small 
$k^2$, this piece contributes an  ${\cal O}\left((\Lambda^2/Q^2)^n\right)$ 
correction from a dimension $n$ condensate. The last integral in 
eq.(2) yields at large $Q^2$  new power
contributions of short distance origin , unrelated to the OPE. If the
short distance power corrections are neglected$^{3)}$ (i.e. if one assumes that
$\delta\alpha_s(k^2)$ is  
sufficiently small  at large
$k^2$), one recovers the standard view$^{5)}$ that the first 
correction to the
Borel sum is given by the OPE. To determine whether this is the case, one needs 
a closer look at $\delta\alpha_s(k^2)$. One may define:
\begin{equation}\delta\alpha_s(k^2)=\delta\alpha_s^{PT}(k^2)+\delta\alpha_s^{NP}
(k^2)
\end{equation}
where $\delta\alpha_s^{NP}$  represents a ``physical'', 
``genuinely non-perturbative''component, which one can assume$^{2)}$ to be 
restricted to low $k^2$: in accordance with the OPE ideology of$^{4)}$, it 
induces
an ${\cal O}\left((\Lambda^2/Q^2)^n\right)$ power correction of IR origin, 
consistent with the OPE, parametrized  with the low energy moment: 
\begin{equation} K_n^{NP} \equiv \int_0^\infty n{dk^2\over k^2}\left({k^2\over
\Lambda^2}
\right)^n
\delta\alpha_s^{NP}(k^2)\end{equation}
(where I extended $\Lambda_I$ to infinity, since the integral is dominated by 
low $k^2$). 
On the other hand, the $\delta\alpha_s^{PT}$ piece is ``unphysical'', 
its role being to remove the Landau pole in $\alpha_s^{PT}$, and has no a
priori 
reason to be restricted to low $k^2$; generically (see below) one expects for 
large $k^2$:
\begin{equation}\delta\alpha_s^{PT}(k^2)\simeq b_{PT}{\Lambda^2\over k^2}
\end{equation}
Besides an (ambiguous) ${\cal O}\left((\Lambda^2/Q^2)^n\right)$ power
correction 
of IR origin, parametrized  with the low 
energy moment: 
\begin{equation} K_n^{PT} \equiv \int_0^{\Lambda_I^2} n{dk^2\over k^2}
\left({k^2\over
\Lambda^2}
\right)^n
\delta\alpha_s^{PT}(k^2)\end{equation}
this piece will induce an (unambiguous) short distance 
${\cal O}\left(\Lambda^2/Q^2\right)$ 
correction, unrelated to the OPE, from the last 
integral in eq.(2). In particular, the range\\ $Q^2<k^2<\infty$ contributes:
\begin{equation}\int_{Q^2}^\infty{dk^2\over k^2}\ \delta\alpha_s^{PT}(k^2)\ 
\varphi\left(k^2
\over Q^2\right)\simeq  A\ 
b_{PT}\ {\Lambda^2\over Q^2}\end{equation}
where $A\equiv\int_{Q^2}^\infty{dk^2\over k^2}\ {Q^2\over k^2}\ \varphi
\left(k^2\over Q^2\right)$ is a number. For instance, the simplest 
``minimal'' regularization of the one loop coupling:
\begin{eqnarray}\alpha_{s,reg}^
{PT}(k^2)&
\equiv&{1\over\beta_0\ln(k^2/\Lambda^2)}- {1\over\beta_0}{1\over{k^2\over
\Lambda^2}-1 }\nonumber\\ &\equiv&\alpha_s^{PT}(k^2) + 
\delta\alpha_s^{PT}(k^2)\end{eqnarray}
gives $b_{PT}=-1/\beta_0$. This  example has the  
interesting 
feature that 
the time-like discontinuity of the regularized coupling coincides with that of 
the perturbative coupling, and suggests a general ansatz (which has actually
been suggested long ago in QED, and has been recently 
revived$^{6)}$ in QCD):
\begin{equation}\alpha_{s,reg}^{PT}(k^2)=k^2\int_0^\infty{d\mu^2\over
(\mu^2+k^2)^2}\
\alpha_{eff}^{PT}(\mu^2)\end{equation}
where the perturbative ``effective coupling'' $\alpha_{eff}^{PT}(\mu^2)$ is 
related to the ``spectral density'' of the perturbative coupling
$\rho_{PT}(\mu^2) \equiv -
\frac{1}
{2\pi i}\{\alpha_s^{PT}\left[-(\mu^2+i\epsilon)\right]-\alpha_s^{PT}
\left[-(\mu^2-i
\epsilon)\right]\}$ by:
\begin{equation} {d\alpha_{eff}^{PT}\over d\ln \mu^2}=\rho_{PT}(\mu^2) 
\end{equation}
The same dispersion relation was assumed in$^{2)}$ for the {\em total} coupling
\begin{equation}\alpha_s=\alpha_{s,reg}^{PT}+\delta\alpha_s^{NP}\end{equation}
i.e.:
\begin{equation}\alpha_s(k^2)=k^2\int_0^\infty{d\mu^2\over(\mu^2+k^2)^2}\
\alpha_{eff}(\mu^2)\end{equation}
where $\alpha_{eff}(\mu^2)$ is related to the discontinuity of $\alpha_s$ by 
eq.(10). Putting:
\begin{equation}\alpha_{eff}(\mu^2)=\alpha_{eff}^{PT}(\mu^2)+\delta
\alpha_{eff}^{NP}
(\mu^2)\end{equation}
it follows that the ``non-perturbative modification'' satifies also:
\begin{equation}\delta\alpha_s^{NP}(k^2)=k^2\int_0^\infty{d\mu^2\over
(\mu^2+k^2)^2}\
\delta\alpha_{eff}^{NP}(\mu^2)\end{equation}
Contrary to $\alpha_s^{PT}(k^2)$,
$\alpha_{eff}^{PT}(\mu^2)$ in eq.(9) is likely to be IR finite, which 
explains$^{7)}$ that
$\alpha_{s,reg}^{PT}(k^2)$ differs from $\alpha_s^{PT}(k^2)$ by power 
corrections. In the one-loop coupling example, one has:
\begin{equation}\alpha_{eff}^{PT}(\mu^2)=\frac{1}{\pi\beta_0}\left[{\pi\over
2}-\arctan\left(\frac{1}{\pi}\ln{\mu^2
\over\Lambda^2}\right)\right]\end{equation}
which is indeed IR finite$^{5)}$:
\begin{equation}\alpha_{eff}^{PT}(\mu^2=0)={1\over \beta_0}=\alpha_{s,reg}^{PT}
(k^2=0)\end{equation}

Corresponding to the split eq.(11), one can distinguish in $D(Q^2)$ a 
``regularized perturbation theory'' piece:
\begin{equation}D_{reg}^{PT}(Q^2)\equiv\int_{0}^\infty{dk^2\over k^2}\ 
\alpha_{s,reg}^{PT}(k^2)\
\varphi\left(k^2\over Q^2\right)\end{equation}
and a ``non-perturbative'' power correction piece:
\begin{equation}\delta D_{NP}(Q^2)=\int_0^\infty{dk^2\over k^2}\
\delta\alpha_s^{NP}(k^2)\ \varphi\left(k^2\over Q^2\right)\end{equation}
Note that $D_{reg}^{PT}$ differs from the Borel sum $D_{PT}$ by ``perturbative''
power corrections:
\begin{equation}\delta D_{PT}(Q^2)=\int_{0}^\infty{dk^2\over k^2}\ 
\delta\alpha_s^{PT}(k^2)\
\varphi\left(k^2\over Q^2\right)\end{equation}
and it is an important issue whether it is possible to disentangle these two 
types of power corrections.

For Minkowskian observables, it is necessary to introduce, instead of 
eq.(1),
a representation in term of $\alpha_{eff}$:
\begin{equation}D(Q^2)=\int_0^\infty{d\mu^2\over\mu^2}\ \alpha_{eff}(\mu^2)\ 
\dot{{\cal
F}}\left({\mu^2\over Q^2}\right)\end{equation}
where ${\cal F}$ is the ``characteristic function''$^{2)}$, i.e. the 
${\cal O}(\alpha_s)$ Feynman diagram computed with a
finite gluon mass$^{5)}$ $\mu^2$, and 
$\dot{{\cal F}}\equiv - d{\cal F}/d\ln \mu^2$. $D_{reg}^{PT}$ is directly 
related to $\alpha_{eff}^{PT}$ by:
\begin{equation}D_{reg}^{PT}(Q^2)=\int_0^\infty{d\mu^2\over\mu^2}\ 
\alpha_{eff}^{PT}(\mu^2)
\ \dot{{\cal
F}}({\mu^2\over Q^2})\end{equation}
Concerning the ``perturbative '' power corrections, I quote the following 
result for an
{\em analytic} small $\mu^2$ behavior of
${\cal F}({\mu^2\over Q^2})$ (in which case the power correction can be shown to
 be of short distance origin). If:
\begin{equation}{\cal F}({\mu^2\over Q^2})-{\cal
F}(0)\simeq -d\ {\mu^2\over Q^2}\ \ \ \ \ \ (\mu^2\ll Q^2)\end{equation}
then:
\begin{equation}\delta D_{PT}(Q^2)\simeq b_{PT}\ d\ 
{\Lambda^2\over Q^2}\ \ \ \ \ \ (Q^2\gg\Lambda^2)\end{equation} 
However, $b_{PT}$ is difficult to calculate, since it depends on the $\alpha_s$
beta-function to {\em all orders} (similarly to IR renormalons residues).

As an application, consider the hadronic width of the $\tau$ lepton. It is 
usually expressed in term of the quantity
$R_{\tau}$, itself related to the total $e^+e^-$ annihilation cross-section 
into hadrons
$R_{e^+e^-}$  by:  
\begin{equation}R_{\tau}(m_{\tau}^2)=2\int_0^{m_{\tau}^2}{ds\over
m_{\tau}^2}\left(1-{s\over m_{\tau}^2}\right)^2 \left(1+2{s\over m_{\tau}^2}
\right)
R_{e^+e^-}(s)\end{equation}
In the small $\mu^2$ limit, one finds$^{2)}$ for the corresponding 
characteristic 
function:
\begin{equation}{\cal F}_{\tau}({\mu^2\over m_{\tau}^2})-{\cal F}_{\tau}(0)
\simeq -d_{\tau}
{\mu^2\over
m_{\tau}^2}\end{equation}
with:
\begin{equation}d_{\tau}={16\over 3\pi}(4-3\zeta(3))\end{equation}
which implies a leading $1/m_{\tau}^2$ power
correction {\em of UV origin}: 
\begin{equation}\delta R_{\tau}^{PT}(m_{\tau}^2)\simeq b_{PT}\ d_{\tau} 
{\Lambda^2\over 
m_{\tau}^2}
\end{equation}
For a numerical estimate, assume the ``large $\beta_0$'' value 
$b_{PT}=-1/\beta_0$, and take:
$\Lambda=\Lambda_V=2.3\Lambda_{\overline {MS}}$ to be the Landau pole of the 
``V-scheme''$^{5)}$ coupling. Then (for 3 flavors):
\begin{equation}\delta R_{\tau}^{PT}(m_{\tau}^2)\simeq -0.934 {\Lambda_V^2\over 
m_{\tau}^2}
\end{equation}
wich gives, assuming e.g.
$\alpha_s^{\overline {MS}}(m_{\tau}^2)=0.32$: $\delta R_{\tau}^{PT}(m_{\tau}^2)
\simeq -0.063$. One thus gets a sizable correction with respect
to the (principal-value) Borel sum estimate$^{5)}$ (still in the large
$\beta_0$ 
limit): $R_{\tau}(m_{\tau}^2)-1\simeq 0.227$ , or to the experimental value:\\
$R_{\tau}(m_{\tau}^2)-1\simeq 0.20$ ($R_{\tau}$ is normalized as $R_{\tau}=1+
{\alpha_s\over \pi}+...$). This result shows that $1/m_{\tau}^2$ terms could 
be at the same level as radiative corrections in $\tau$ decay (where standard 
power contributions of IR origin are estimated to be very small!). Note
also that 
a corresponding $1/Q^2$ power correction is absent from $R_{e^+e^-}^{PT}(Q^2)$
(for which the leading power correction (of UV origin) is only$^{2)}$ ${\cal O}
(1/Q^4)$).

In conclusion, the removal of the Landau singularity is likely to induce power 
corrections of UV origin (hence unrelated to, thus not inconsistent with, the
OPE), which a priori should be of similar size as higher order radiative 
corrections. An important issue is to assess 
whether these corrections will modify in a significant way the standard IR power
contributions
phenomenology. As a first guess, one might expect them to be relevant in 
processes where many orders of perturbation theory should be taken into account,
such as inclusive $\tau$ decay, or to handle the ``perturbative tail'' of the 
gluon condensate on the lattice.

Similar remarks have recently been  put forward by R. Akhoury and V.I. Zakharov 
(hep-ph/9705318).\\

%
\noindent
{\bf References}\\

\noindent 1) G. Grunberg, hep-ph/9705290.\\

\noindent 2) Yu.L. Dokshitzer, G. Marchesini and B.R. Webber,  {\sl Nucl. Phys.}
 {\bf B469} 
(1996) 93.\\

\noindent 3) Yu.L. Dokshitzer and B.R. Webber,  {\sl Phys.Lett.} {\bf B352} 
(1995) 451.\\

\noindent 4) M.A. Shifman, A.I. Vainshtein and V.I. Zakharov, 
{\sl Nucl. Phys.} {\bf B147}  
(1979) 385.\\

\noindent 5)  
P. Ball, M. Beneke
and V.M. Braun, {\sl Nucl. Phys.} {\bf B452} (1995) 563.\\

\noindent 6) D.V. Shirkov and I.L. Solovtsov, hep-ph/9704333; A.I Alekseev and
 B.A. Arbuzov, 
hep-ph/9704228 and references therein.\\

\noindent 7) G. Grunberg,  {\sl Phys.Lett.} {\bf B372} (1996) 121; ibid. 
hep-ph/9608375; Yu.L. Dokshitzer and N.G. Uraltsev, {\sl Phys.Lett.} {\bf B380}
(1996) 141.\\

\end{document}